# A Neural Network Approach to Predict Gibbs Free Energy of Ternary Solid Solutions

Paul Laiu[1], Ying Yang[2], Massimiliano Lupo Pasini[3], Jong Youl Choi[1], Dongwon Shin[2*]

[1]*Computer Science and Mathematics Division, Oak Ridge National Laboratory, Oak Ridge, TN 37831*

[2]*Materials Science and Technology Division, Oak Ridge National Laboratory, Oak Ridge, TN 37831*

[3]*Computational Sciences and Engineering Division, Oak Ridge National Laboratory, Oak Ridge, TN 37831*

Abstract

We present a data-centric deep learning (DL) approach using neural networks (NNs) to predict the thermodynamics of ternary solid solutions. We explore how NNs can be trained with a dataset of Gibbs free energies computed from a CALPHAD database to predict ternary systems as a function of composition and temperature. We have chosen the energetics of the FCC solid solution phase in 226 binaries consisting of 23 elements at 11 different temperatures to demonstrate the feasibility. The number of binary data points included in the present study is 102,000. We select six ternaries to augment the binary dataset to investigate their influence on the NN prediction accuracy. We examine the sensitivity of data sampling on the prediction accuracy of NNs over selected ternary systems. It is anticipated that the current DL workflow can be further elevated by integrating advanced descriptors beyond the elemental composition and more curated training datasets to improve prediction accuracy and applicability.

---

[*] Corresponding author: D. Shin (shind@ornl.gov)

# 1. Introduction

Phases are critical microstructural features that govern material properties. A presentative example is the superior high-temperature mechanical properties of Ni-based superalloys primarily benefit from the formation of the Ni-based FCC phase and $Ni_3Al$-based $L1_2$ phase [1]. Since state-of-the-art structural alloys are often made from many alloying elements, and accurately predicting the phase stability knowledge in the high-dimensional multi-element space is challenging. The CALPHAD (CALculation of PHAse Diagram) approach [2] is the only viable way to calculate multi-component phase diagrams at finite-temperatures. The accuracy of the computed phase equilibria from the CALPHAD approach depends on the reliability of the Gibbs energy functions of individual phases. Therefore, modeling the Gibbs energy function of individual phases is the core component of the CALPHAD approach.

The Gibbs energy of individual phases is modeled as a function of composition and temperature based on crystal structure, defect type, and phase chemistry. The model parameters are then evaluated through an optimization procedure that aims to self-consistently reproduce the experimentally assessed phase equilibria and thermodynamic properties. As more experimental data are available for lower-order systems (i.e., unary and binary) than higher-order systems (i.e., ternary, quaternary, and beyond), Gibbs energy functions of a phase in higher-order systems were constructed through extrapolation of Gibbs energy functions modeled for phases in lower-order systems. Typically, the Gibbs energy functions developed for single-principal-element alloys include Gibbs energy functions of phases in those binary and ternary systems containing the principal element and are only valid for the principal-element-rich region.

This strategy worked well for single-principal-element alloys because the Gibbs energy contribution for a phase from the ternary and/or higher-order interaction becomes small when those interaction parameters are multiplied by the small concentrations of minor elements [3]. However, the Gibbs energy function of a ternary or higher-order phase developed in such a way is not adequate for making accurate predictions for high-entropy alloys where all elements are, in principle, equally important. Incorporating additional ternary or higher-order interactional parameters is necessary through optimization based on experimental data. For *n*-component phases, the number of constituent systems with binary interaction is $C_n^2$, and that for systems with ternary interaction is $C_n^3$. The total number of ternary systems to be modeled increases significantly with the growing number of components [4]. The high dimensionality and the lack of ternary experimental data hamper the complete thermodynamic modeling of all constituent ternary systems.

Machine learning has been increasingly coupled with density functional theory (DFT) and experimental data to predict various thermodynamic properties such as Gibbs free energy, formation enthalpies/entropies, and heat capacity [5–9]. However, the existing studies were focused on stoichiometric compounds with very well-defined chemical ordering and chemistry at zero Kelvin. To the best of the authors' knowledge, no data analytics work has been done on predicting the energetics of solid solution phases with a wide range of compositions at finite-temperatures. The difficulty is primarily due to the limited capability of DFT calculation and the lack of experimental data in the Gibbs energy of solid solution phases to generate large datasets for training surrogate models.

Herein, we present a data-centric computational workflow based on an emerging deep learning (DL) approach using neural networks (NNs) to demonstrate the feasibility of predicting Gibbs free energies of solid solutions. We explore how to train NNs with a dataset of Gibbs free energies computed from a state-of-the-art CALPHAD database to predict ternary systems. We investigate the sensitivity of the prediction accuracy of NNs with respect to data sampling over several selected ternary systems. We then briefly discuss the limitations of the current DL approach. Finally, we propose a future research direction to further improve the accuracy and applicability of the DL approach to predicting energetics of ternary solid solutions.

## 2. Computational Approach

We explore recent data-driven NN methods for high-order prediction in the present paper. The application of NNs, also widely known as artificial neural networks or simulated neural networks, has shown success in many scientific domains, including the materials sciences [10–19]. Under the inspiration of the networks of neurons in the human brain, NNs are comprised of layers of neuron nodes, including an input layer, one or more hidden layers, and an output layer. Each node, or artificial neuron, connects to another and has an associated weight and threshold, also known as the activation function. We optimize the weights and threshold parameters of nodes during the training process by using a user-defined loss function with a dataset, called a training set. After completing training, we use the trained NNs for prediction or classification with the dataset not used during the training, known as validation.

The overall approach of the present study is illustrated in Figure 1. We start with computing Gibbs free energy of binaries and ternaries from the CALPHAD database to construct a dataset for NN training. We train two NN models with different datasets: i) one with only Gibbs free energies of binaries and ii) the other augmented with a small set of select ternaries in addition to the binary dataset. Our main objective is to compare NN predictions trained with and without augmented ternary data and propose an efficient NN training method with minimal augmented ternary data for better prediction accuracy and generalizability. The performance of the two NN surrogate models in predicting Gibbs free energies of ternary systems has been evaluated by comparing the NN predictions and ground truth computed from the critically assessed CALPHAD database.

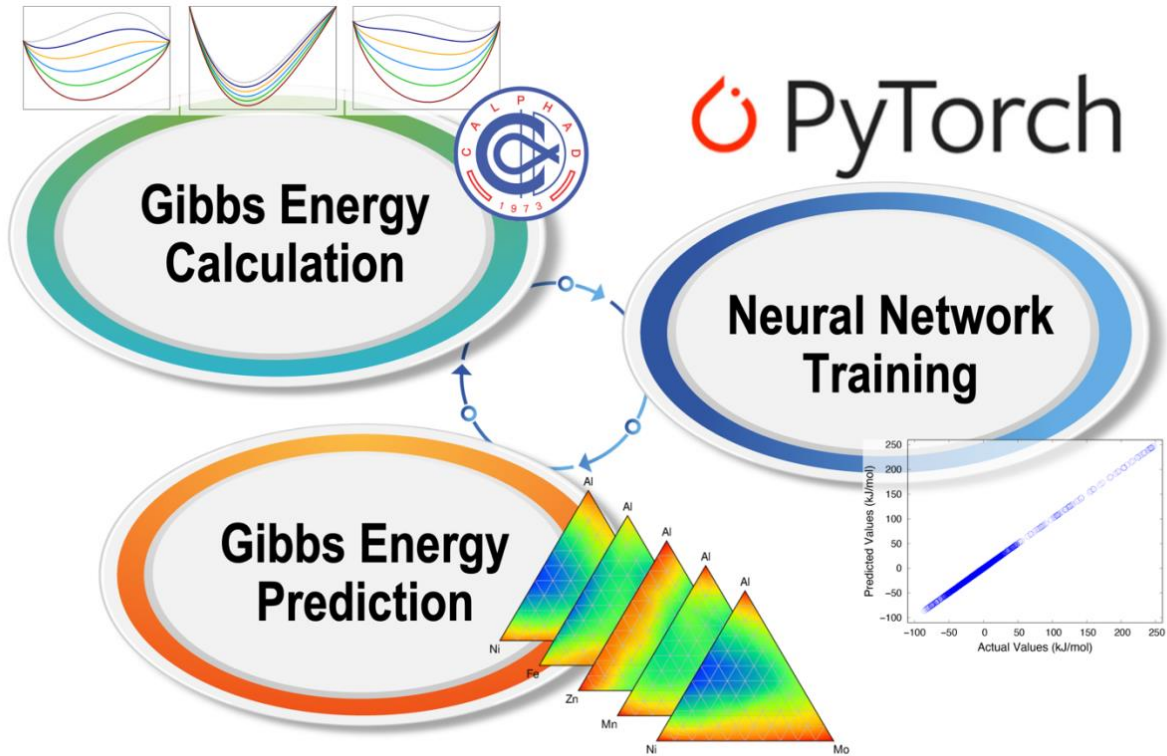

Figure 1: Schematic diagram of the proposed approach

2.1. Computation of the training dataset

We employ a high-throughput CALPHAD approach to rapidly populate the training database for NN training. We have used the high-entropy alloy database v5 (TCHEA5) as implemented in the Thermo-Calc software package. We select the face-centered cubic (FCC) as a test case to demonstrate the feasibility of applying NN surrogate models to predict Gibbs free energies of solid solutions. Twenty-three elements and 226 binaries are considered at 11 different temperatures, i.e., 300K to 1300K, with a uniform increment of 100K. After reviewing the calculated datasets, the following binaries, which were not critically evaluated in the TCHEA5 database, are excluded from training: Ir-Y, Mo-Sn, Nb-Rh, Rh-Si, Rh-Ta, Rh-V, and Rh-Y. Another group of binaries is also excluded because the mixing enthalpies of these systems are modeled as ideal, which should not be: Al-Rh, Hf-Si, Hf-Zn, Ir-Sn, Ir-Zn, Mo-Y, Nb-Si, Nb-Sn, Re-Rh, Re-Zn, Si-Sn, Si-Ta, Si-Y, Si-Zn, Si-Zr, Sn-W, Ta-W, Ta-Zn, Ti-Zn, and W-Zn. The number of binary data points included in the present study is 102,000. We also consider a certain amount of data points from ternaries. Out of 1,771 possible ternaries in 23 element systems, we select six, i.e., Al-Cr-Ni, Al-Cu-Ni, Cr-Fe-Ni, Al-Cu-Fe, Fe-Nb-Ni, and Co-Cu-Mn, to investigate their influence on the prediction accuracy.

2.2. Neural network training

We use NN surrogate models to predict the Gibbs free energy as a function of element compositions and temperatures using the PyTorch framework [20]. Specifically, we adopt the residual neural network (ResNet) architecture [21]. ResNet optimizes residual mapping between layers using skip connections that bypass particular intermediate layers. It shows the ability to avoid the issues of vanishing gradients and saturating accuracy in the training process with many layers. Thus, it improves the stability and accuracy over the standard fully connected multi-layer perceptron. In the ResNet implementations considered in this work, we use the architecture of an input layer, 12 hidden layers, and an output layer. The numbers of nodes in the input and output layers are 24 (23 elements + temperature) and 1 (Gibbs free energy), respectively. The 12 hidden layers consist of four layers with 32 nodes, four layers with 16 nodes, and four layers with eight nodes, with double-layer skip connections in-between. The Rectified Linear Unit (ReLu) activation function is used in each layer except for the output layer, in which a linear activation function allows for negative outputs.

Before training ResNets, we follow the standard practice and pre-process the input and output data by shifting and linearly scaling the temperatures and Gibbs free energies to the intervals [0, 1] and [-1, 1], respectively. We do not pre-process or rescale the composition data as it naturally lies in [0, 1] and has important column-wise (i.e., element-wise) correlations to be maintained. We apply the common 80-20 training-validation split to data for binary systems via random drawing under a uniform probability distribution, whereas all data for unary systems are included in the training dataset. As for the ternary data augmentation tests considered in Sections 3.3 and 3.4, the ternary data are augmented only into the training dataset, with the validation dataset remaining unchanged. The loss function used in the present study is formulated using the standard mean squared error (MSE) on the output, which is minimized during the training process using the Adam optimizer [22]. The learning rate is initialized to 0.001 and refined by a factor of 10 using the `ReduceLROnPlateau` scheduler with patience set to 100. The training process terminates either when it takes the maximum 2000 epochs or when no improvement was observed in the validation error in the last 300 epochs.

## 3. Results

### 3.1. Neural network training results

For all NNs considered in this work, we observe that training and validation errors reached around $10^{-6}$ in MSE after the training step, which translates to a 0.1% network prediction error for both the training and validation datasets. Figure 2 shows a parity plot that compares the actual data to predictions from an example network and the loss curves that track the training and validation errors as the number of epochs grows in the training process. The sign of overfitting, which would manifest itself as a much lower training error than the validation error, is not observed in Figure 2. This result confirms the effectiveness of the choice of the small ResNet architecture discussed in Section 2.2.

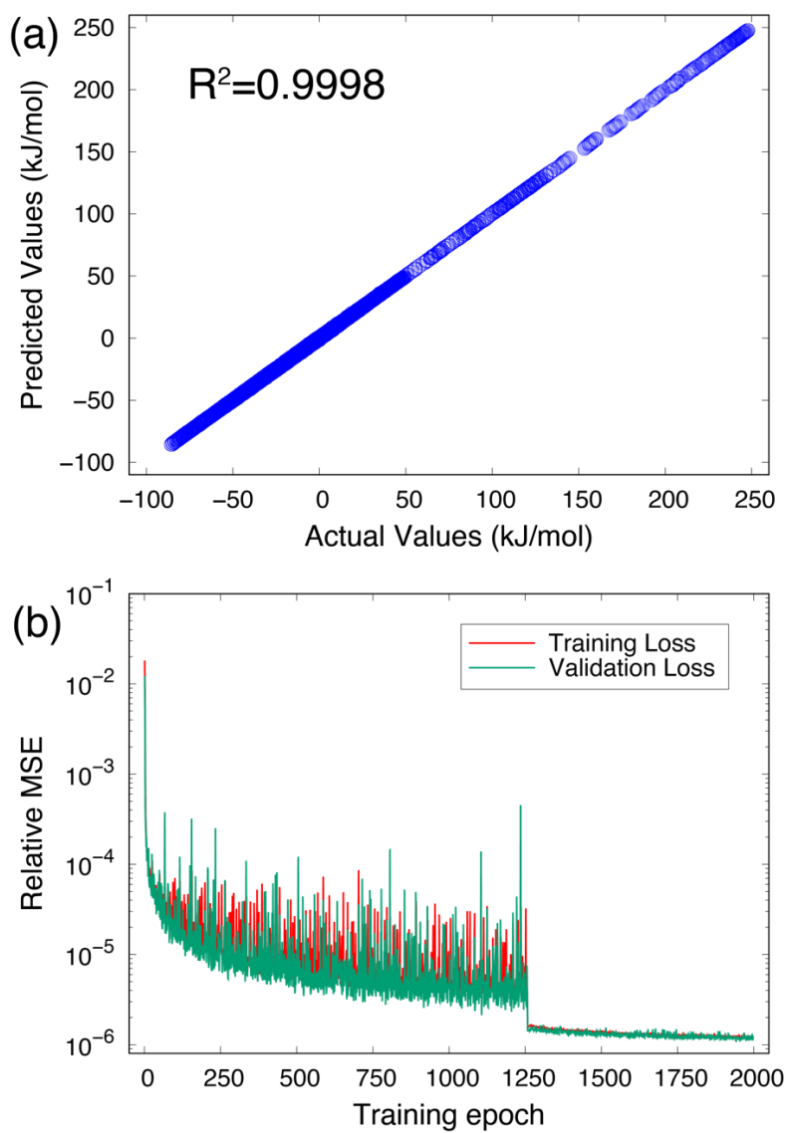

Figure 2: Parity plot and training progress of an example network

3.2. Prediction only with a binary dataset

It is possible to make predictions in ternaries from the NN trained with only the binary Gibbs free energies of FCC solid solutions. We started with making predictions in the select six ternaries, i.e., Al-Cr-Ni, Al-Cu-Ni, Cr-Fe-Ni, Al-Cu-Fe, Fe-Nb-Ni, and Co-Cu-Mn. Figure 3 shows the predicted Gibbs free energies of Al-Cr-Ni and Al-Cu-Fe isopleths at three iso-concentration lines at multiple temperatures (circles) compared with the computed values from the TCHEA5 CALPHAD database (lines). The same plots for the other four ternaries are provided in the supplementary material. While the predicted magnitude is slightly off, it is encouraging that the overall trend in the curvature and temperature sequence has been well replicated in both Al-Cu-Ni and Al-Cu-Fe, which is the case for the other four ternaries as well (see Figures S1 and S2). It should be noted here that NN predicted Al-Cu-Ni to have miscibility gaps in Ni-Al/Cu, as shown in Figure 3(a). The version of the Al-Cu binary model within the TCHEA5 has a small miscibility gap, which can be regarded as a minor artifact during the modeling. It appears that the NN model candidly learned from the constituent binaries.

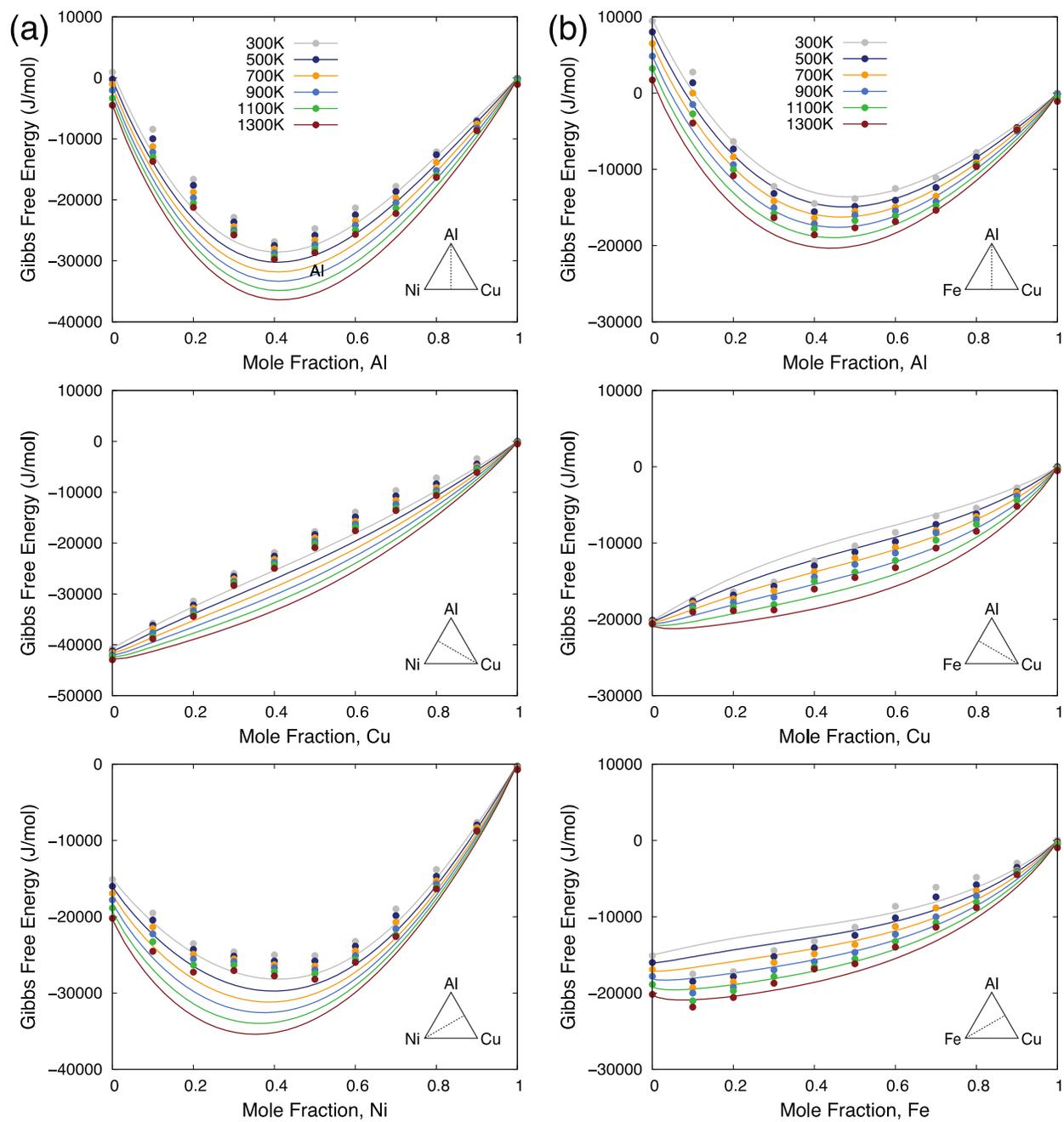

Figure 3: Isopleths of Al-Cu-Ni and Al-Cu-Fe at multiple temperatures

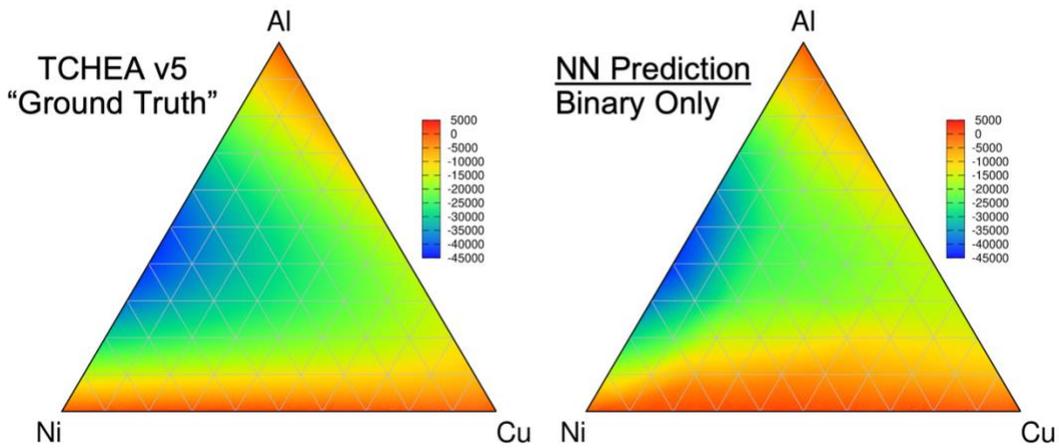

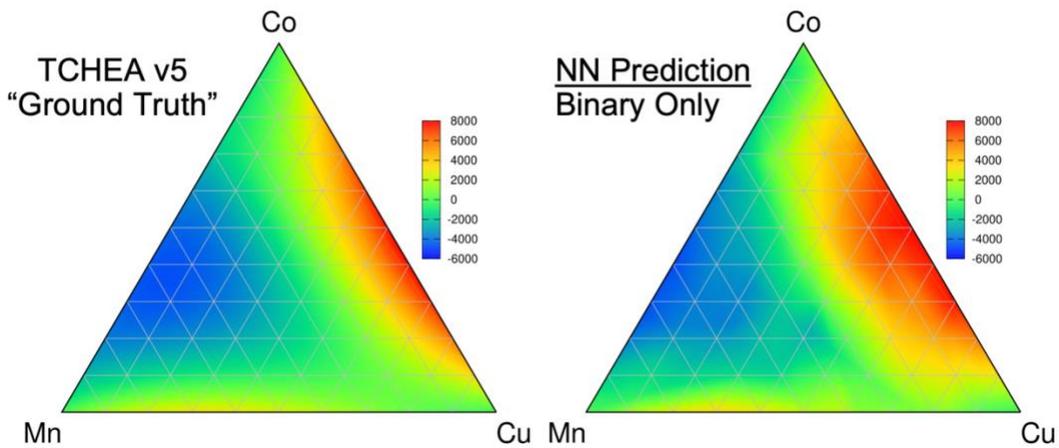

Figure 4: Ternary projection of Al-Cu-Ni and Co-Cu-Mn at 300K

We also compared the predictions from the NN trained with only binary data to the computed results from the TCHEA5 database at 300K, presented in Figure 4. Similar to the comparison in the isopleth plots, the NN-predicted and computed Gibbs free energy contour plots at 300K overall show a good agreement for Al-Cu-Ni and Co-Cu-Mn. While the change in the energy within the Gibbs triangles of NN-predicted values is not as smooth as the computed values, the respective minimum and maximum areas are well-reproduced. The comparison between NN-predicted and computed Gibbs free energies of the other four ternaries is provided in the supplementary material (see Figures S3 and S4), and they also show a similar trend.

The materials science community widely accepts that the most critical interactions in multi-component systems are those of binaries. In many cases, a good approximation of ternary energetics can be obtained without introducing ternary interactions. On the other hand, interpolated energies from constituent binaries are insufficient to adequately describe ternary energetics, and ternary interactions are required in some instances. Figure 5a compares the computed Gibbs free energetics of the FCC solid solution phase in Co-Cu-Mn with and without ternary interaction parameters at 300K. It is clearly shown that ternary interaction parameters were introduced to adjust Gibbs free energies to be more negative on the Co-Mn side and less positive on the Co-Cu side. Figure 5b shows that assessed ternary Gibbs free energies of the FCC phase in Co-Cu-Mn at 700 and 900K are more negative than those interpolated from constituent binaries.

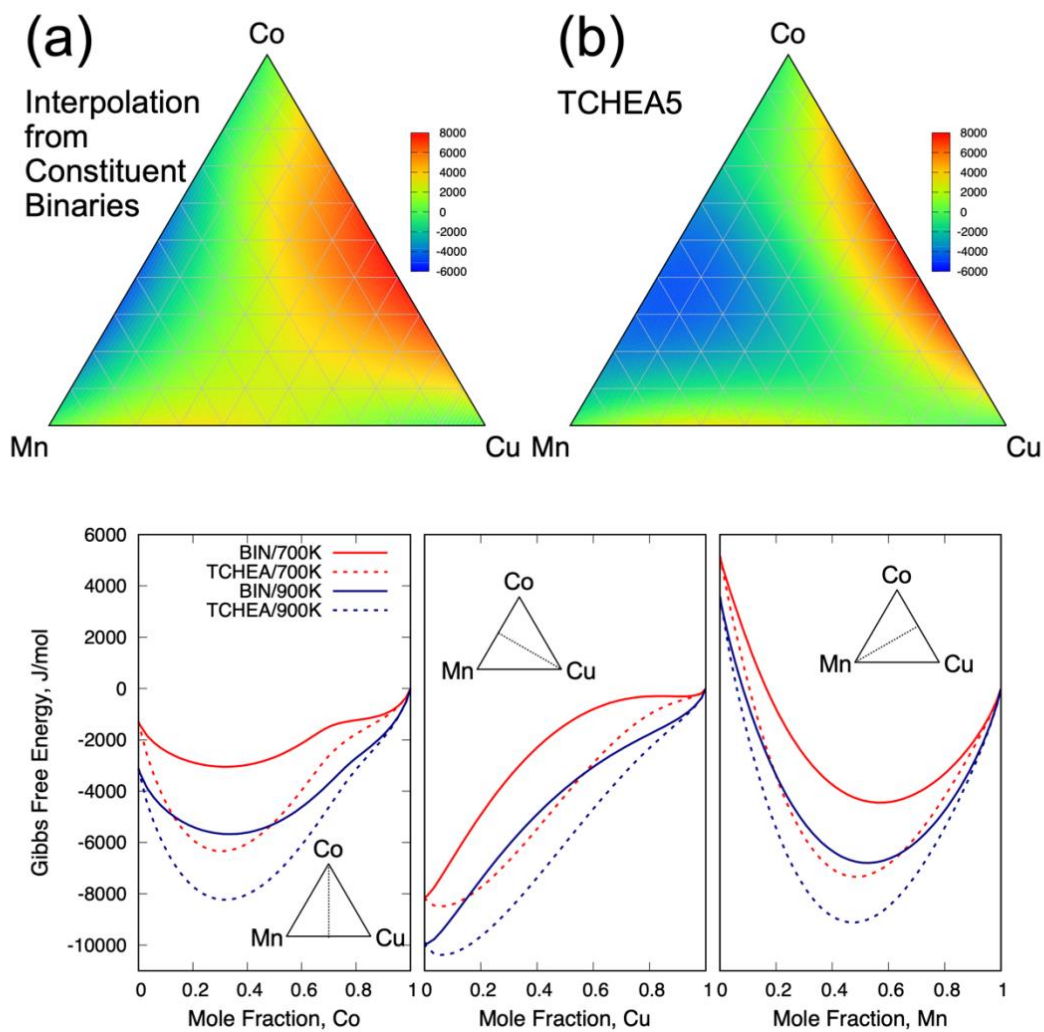

Figure 5: Gibbs free energy of Co-Cu-Mn — binary interpolation vs. assessed ternaries

It would be ideal that an NN could heuristically infer this ternary interaction by learning only from a large number of binaries. From these observations, the Gibbs free energies predicted from the NN trained only with binaries show good qualitative agreement; however, the discrepancy between the NN-predicted and computed ground truth values is somewhat large. Thus, using the NN predicted results as they are, would not be possible where high accuracy is required, e.g., evaluating Gibbs energies for thermodynamic assessment. It motivates our next experiment to use mixed datasets with binaries and ternaries.

3.3. Training of neural network model with ternary data augmentation

While it is shown in Section 3.2 that the NN surrogate model only trained with many binaries can make encouraging predictions that qualitatively replicate trends, there is a considerable discrepancy to the computed ground truth values from the assessed thermodynamic models. In addition, some ternaries indeed require ternary interaction parameters. For these reasons, we investigate the effect of augmenting ternary data into the training dataset on the accuracy of NN predictions. There are 1,771 ternaries in the 23 elements multi-component system. Only a small fraction of ternaries has been critically assessed due to the high dimensionality and the lack of reliable experimental data. Thus, we have carefully selected six ternaries, i.e., Al-Cr-Ni, Al-Cu-Ni, Cr-Fe-Ni, Al-Cu-Fe, Fe-Nb-Ni, and Co-Cu-Mn, and computed Gibbs free energies of FCC at 11 temperatures from 300K to 1100K with a uniform increment of 100K.

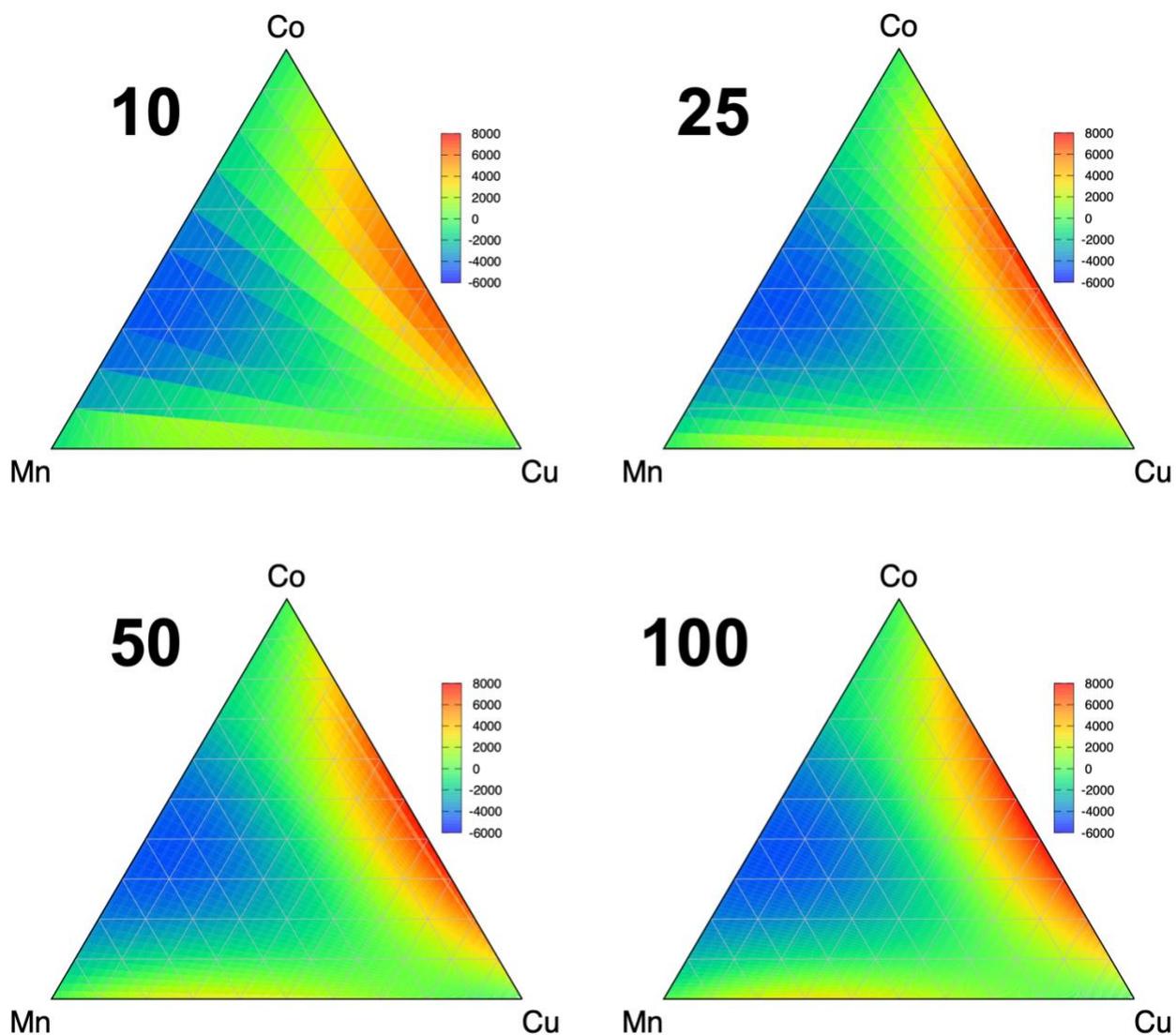

Figure 6: Gibbs free energy of Co-Cu-Mn at 300K with varying number of isopleths

We have investigated the data density by changing the number of isopleths in ternary to prevent the dataset from becoming highly biased toward the ternary over binaries, as illustrated in Figure 6. We tested 9, 24, 49, and 99 isopleths as shown in Co-Cu-Mn at 300K. As the number of isopleths increases, i.e., the resolution gets higher, more details can be captured; however, the volume of data also increases dramatically. We chose 24 isopleths in each temperature and ternary system with ten data points in each isopleth to collect initial data.

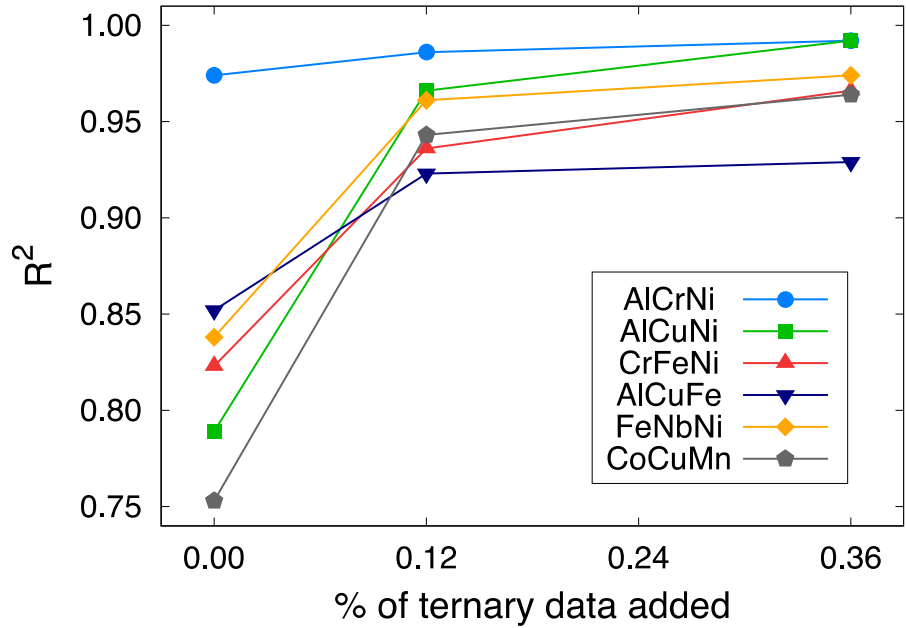

Figure 7: $R^2$ values of NN prediction versus additional ternary data volumes

We then investigated the improvement of NN prediction accuracy with the varying amount of ternary data used to train NN. Figure 7 shows the accuracy of NN, represented with the coefficient of determination ($R^2$) as a function of the amount of augmented ternary data. As anticipated, the accuracy of six ternary predictions improves as the number of ternary data increases. Intriguingly, we also observe that the accuracy improvement from adding ternary data diminished when more than 0.36% of the ternary data were added. We conjecture the diminishing improvement reflects the limitation of the current NN architecture. Using a more complicated NN architecture could potentially lead to further accuracy improvement with additional ternary data augmentation, at the risk of overfitting.

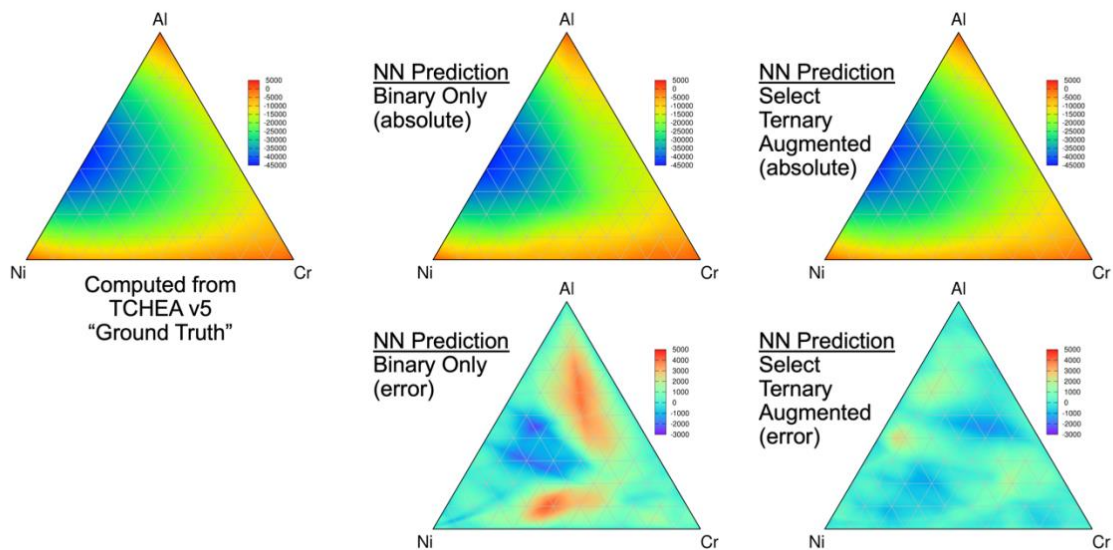

Figure 8: Predicted Gibbs free energy of Al-Cr-Ni at 850K

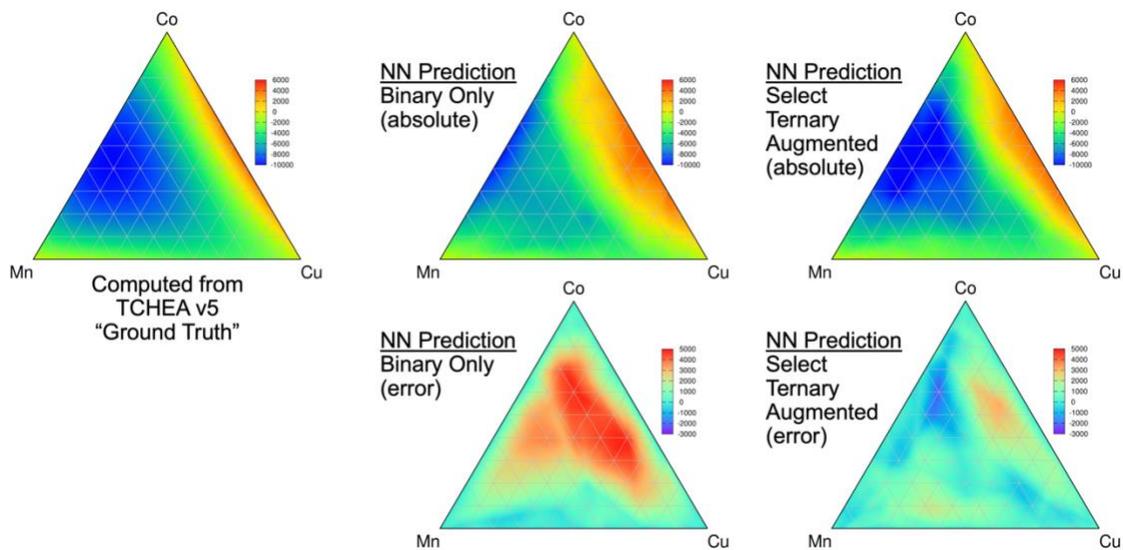

Figure 9: Predicted Gibbs free energy of Co-Cu-Mn at 850K

We then investigate the predictive capability and evaluate errors of NN surrogate models trained with two different datasets, i.e., only with binary energies and augmented with six selected ternaries. Figures 8 and 9 show the results of Al-Cr-Ni and Co-Cu-Mn at 850K, a temperature not included in the training dataset. Figure 8 clearly shows that the NN model trained only with the binary dataset can make a good overall prediction in Al-Cr-Ni. Still, the NN model trained with an augmented ternary dataset further improved the prediction. The evaluated error between the ground truth and NN prediction with ternary augmented dataset is notably smaller than that of prediction from NN trained only with the binary dataset. The same prediction and error evaluation for Co-Cu-Mn presented in Figure 9 shows a similar trend. In Co-Cu-Mn, the prediction from the NN model trained only with binary dataset overpredicted the Gibbs free energies in the Co-Cu side. These errors were much reduced in the NN prediction with the ternary augmented dataset. The results of the other four ternaries are provided in the supplementary material (Figures S5-S8), and they commonly follow the trend shown in Al-Cr-Ni and Co-Cu-Mn: the error gets smaller in the predictions from NN models trained with the augmented ternary dataset. Thus, it can be tentatively concluded that including a ternary dataset in NN training improves prediction accuracy, even with a small amount of additional data.

3.4. 'Extrapolation' with augmented ternary data

It is encouraging that including a small amount of ternary dataset positively influences the accuracy of ternary prediction. However, the case for the six ternaries presented in Figures 8 and 9 (as well as Figures S5-S8 in the supplementary material) is more validating self-correction than making pure predictions. Thus, we investigate the performance of NNs trained with and without the augmented ternary dataset in predicting ternary data that have not been augmented. We have considered two different cases: 1) all the three elements in a ternary are constituents of six ternaries that we have augmented, and 2) one of the elements is outside the six ternaries. The composition of the 23 elements defines a simplex in the 23-dimensional space. Any multi-component alloy is identified at a given temperature by a point inside this simplex. All possible ways to combine three elements span a triangle in this simplex, where the corners/points represent pure elements and the sides (i.e., lines) represent binary alloys. Since each ternary alloy can be expressed as a linear convex combination between two pairs of points that sit on the corners and sides of the triangle, our DL approach does not extrapolate beyond the range of the given training data. Therefore, the Gibbs free energy prediction for a given ternary alloy does not indicate that the NN is extrapolating outside the training region.

### 3.4.1. Constituent elements within six ternaries

There are eight constituent elements, i.e., Al, Co, Cr, Cu, Fe, Mn, Nb, and Ni, in the six ternaries we have generated for augmentation. We selected two subset ternaries, Al-Cu-Mn and Al-Fe-Mn, of which ternary data were not used for augmentation as test cases. It can be anticipated that the NN surrogate model trained with an augmented ternary dataset may perform better than the one only trained with binaries, as the former is expected to learn more about energy landscape from other ternary data. Figures 10 and 11 show the predicted Gibbs free energy of FCC in Al-Cu-Mn and Al-Fe-Mn at 850K from NN models trained with binary-only and with the augmented ternary dataset, respectively, with the ground truth data as well as evaluated errors.

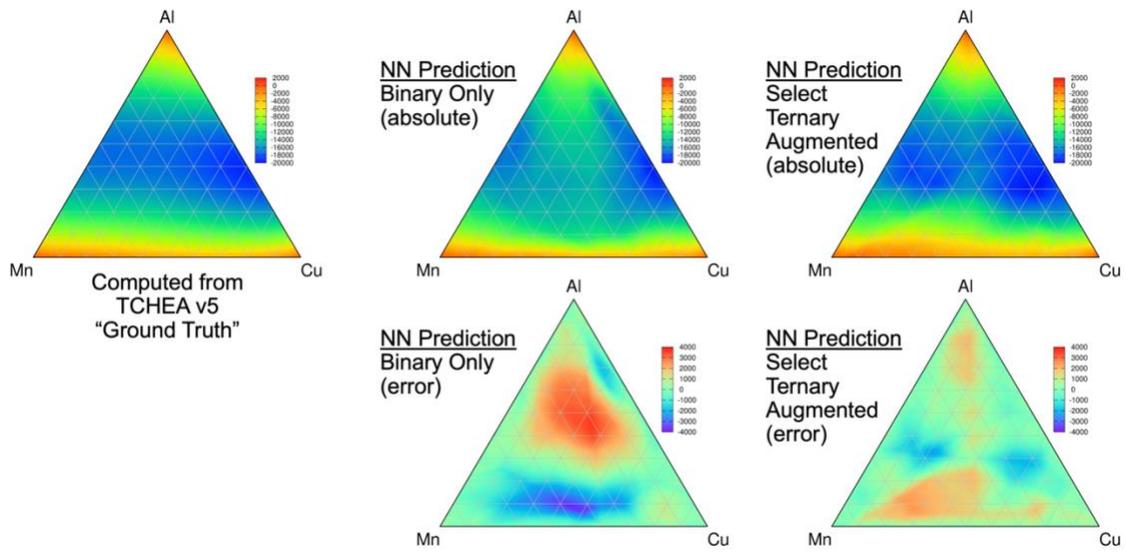

Figure 10: Predicted Gibbs free energy of Al-Cu-Mn at 850K

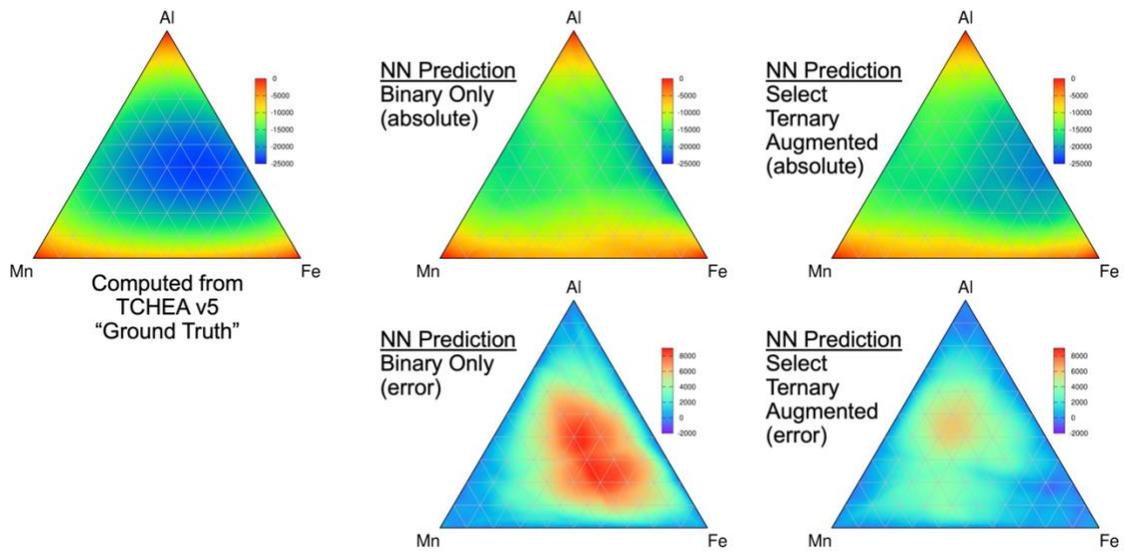

Figure 11: Predicted Gibbs free energy of Al-Fe-Mn at 850K

As anticipated, the NN model trained with augmented ternary data performs better than the one only with binary in both Al-Cu-Mn and Al-Fe-Mn, with overall lower errors. In Figure 10, both NN models reproduced the trend well overall; however, the NN model trained with binary-only data overpredicted the area from the Al-rich corner to the center and slightly underpredicted the Mn-Cu side. The NN prediction with selected ternary augmented dataset successfully captured continuously negative energies between Al-Mn and Al-Cu, and the error is smaller than its binary-only counterpart. A similar observation is made for Al-Fe-Mn that the NN model trained with ternary dataset outperforms the one trained only with binaries. There is a large overpredicted area from the center to the Al-Fe side in the prediction from NN trained only with the binary dataset. Intriguingly, both NN predictions did not successfully capture the largely negative energies in the middle.

*3.4.2. Elements outside of six ternaries for augmentation*

Next, we extend our NN predictions to ternaries with a single element outside of the eight elements in the six ternaries for augmentation. We selected Zn and Mo and made NN predictions of Al-Cu-Zn and Al-Mo-Ni. Figure 12 shows the predicted Gibbs free energy of Al-Cu-Zn at 850K. The same trend was observed in the NN trained with ternary augmented dataset performance better than the one trained only with binaries. In both cases, the error is predominant in the Zn-rich corner. As shown in Figure 13, the results for Al-Mo-Ni are also interesting. While both NN predictions qualitatively reproduced the Gibbs energy landscape of Al-Mo-Ni at 850K, the error on the NN prediction trained with an augmented ternary dataset is slightly worse than the one trained only with binary.

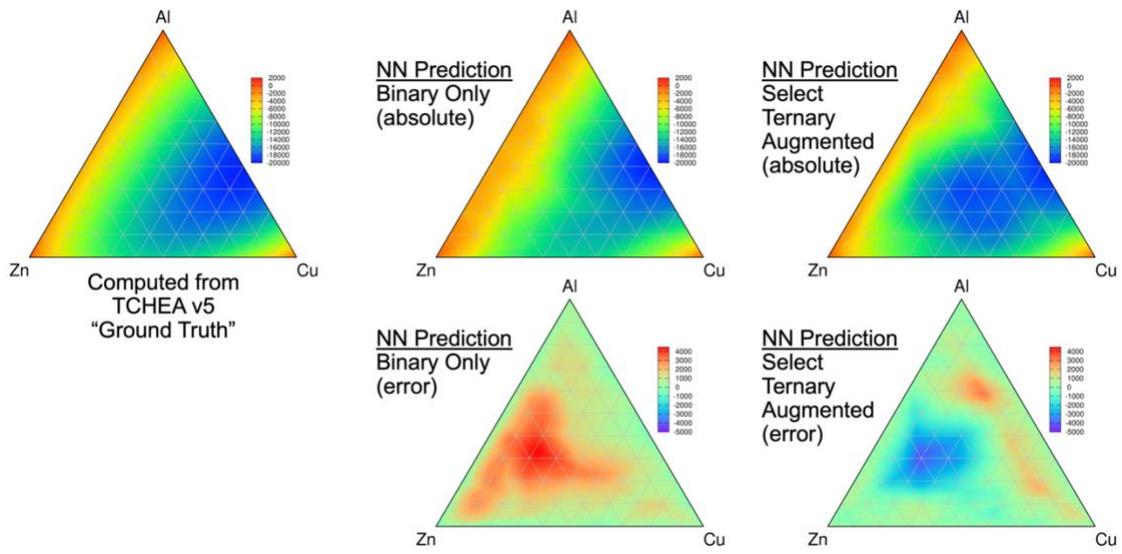

Figure 12: Predicted Gibbs free energy of Al-Cu-Zn at 850K

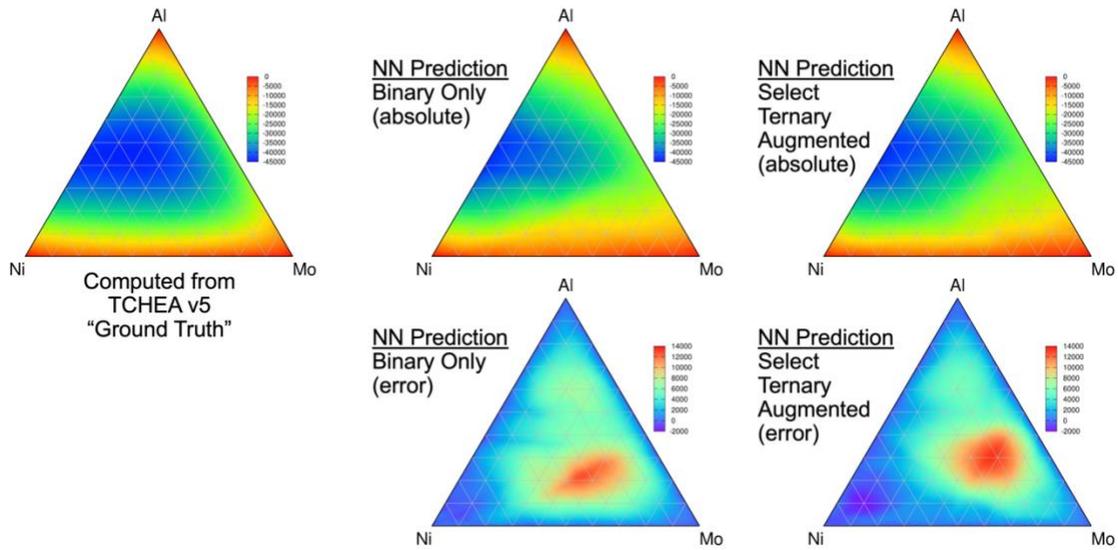

Figure 13: Predicted Gibbs free energy of Al-Mo-Ni at 850K

## 4. Discussion

### 4.1. The choice of a dataset for neural network training

Data science methodologies are split between two categories: data-centric and model-centric. The former assumes that a fixed statistical model has sufficient predictive power to attain the desired accuracy contingent on an accurate collection of data for training. The latter assumes that the data is already well representative of the sampling space and focuses on changing the statistical model to enhance its attainable accuracy. As demonstrated throughout the present study, the methodology employed in the current work is highly data-centric. We describe how the choice of the training dataset directly influences the performance of NN prediction.

The performance of a NN trained only with the binary Gibbs free energies was encouraging; however, predicted results on ternary systems were not good enough compared to the ground truth data computed from the state-of-the-art CALPHAD database, even with the nearly perfect $R^2$ value for binary predictions (see Figure 2). The discrepancy with the prediction from the NN only with the binary data and the ground truth could be attributed to the overlooked ternary interaction, which cannot be simply interpolated from the binary interpolation. The NN surrogate model trained with augmented ternary exhibited better performance than the one trained only with the binary dataset.

A series of ternary predictions with the NNs have shown that the performance of surrogate models is susceptible to the specific ternaries included in the augmentation. For example, the performance of NN surrogate models trained with augmented six ternaries (i.e., Al-Cr-Ni, Al-Cu-Ni, Cr-Fe-Ni, Al-Cu-Fe, Fe-Nb-Ni, and Co-Cu-Mn) was better on predicting Al-Cu-Mn than Al-Fe-Mn. The difference in predictive performance can be explained by the more available learning data for Al-Cu-Mn than Al-Fe-Mn. In particular, the NN surrogate models had a chance to learn about ternary energetics of Al-Cu-Mn with the constituent binaries (i.e., Al-Cu and Cu-Mn) from three ternaries within six: Al-Cu-Ni, Al-Cu-Fe, and Co-Cu-Mn. Thus, the NN network has more information regarding Al-Cu-X (X=Ni and Fe) and Cu-Mn-X (X=Co) to make predictions for Al-Cu-Mn.

On the contrary, there is only one relevant ternary system out of six for Al-Fe-Mn, i.e., Al-Cu-Fe. Thus, NN only had a chance to learn about Al-Fe-X (X=Cu), and no ternaries about constituent binaries, i.e., Al-Mn or Fe-Mn, can be learned. As shown in the case for Al-Cu-Zn and Al-Mo-Ni examples, NN overall works satisfactorily; however, NN surrogate models struggle to make accurate predictions in systems with insufficient data. This is because we have used a single global fitting model that covers the entire simplex instead of the local polynomial fitting models currently employed in the CALPHAD approach. The training of the model aims to identify a compromise between the accuracy of different portions of the data, which explains why improving the accuracy in some regions of the simplex may slightly compromise the accuracy in other areas.

In addition, the difference in the NN predictions of Al-Cu-Zn (good) and Al-Mo-Ni (bad) can be explained by the difference in the stable FCC region in Al-Zn and Al-Mo. While Al-Zn has an extensive stable FCC single-phase region, Al-Mo has a very narrow FCC region. Thus, we can regard the modeling of the FCC phase in Al-Mo as an extrapolation. Therefore, it can be postulated that experimental data and interaction in the ternary assessment compensated for the error resulting from the Al-Mo side. Since our NN did not see any ternary data of Al-Mo-Ni, it produces a significant discrepancy. On the contrary, the assessed Gibbs energy of FCC in Al-Zn can be considered a reliable ground truth because of the large stable region. Similarly, other constituent binaries, i.e., Al-Cu and Cu-Zn, also have large single FCC regions. Thus, both CALPHAD and NN have enough data to learn from constituent binaries and show better agreement. Therefore, it can be concluded the accuracy of the NN surrogate model to predict Gibbs free energy of ternary solid solution is directly related to the quantity, quality, and statistical sampling of the dataset used within.

As a data-driven surrogate model, the performance of the NN is highly sensitive to the volume and variety of the dataset. In other words, it is anticipated that the behavior of NN prediction is highly biased toward data used within the training. This principle is true for many physical and analytical model development processes. The importance of critically evaluating available experimental/theoretical datasets was already pointed out in the CALPHAD community [23]. It was also emphasized that bad and contradicting data should be eliminated prior to the model development procedure, which requires considerable domain knowledge.

The grand challenge in the demonstrated data analytics framework in the present study to predict Gibbs free energies would be selecting ternary systems for augmentation. We have chosen six ternaries to investigate their effect on NN prediction based on the assumption that these ternary data can be used as ground truth for NN training. However, it would be very challenging to validate that these synthetic data computed from the CALPHAD model can be regarded as 'ground truth' due to the difficulties in experimental measurements and the accuracy of theoretical calculations, such as DFT [24–26]. It is the same reason why interaction parameters of solid solution phases for most ternary systems remain untouched.

On the positive side, a certain number of ternary systems have been critically evaluated, and ternary interaction parameters for solid solutions have been confidently introduced. Thus, the first step to exploiting the emerging NN approach to predict Gibbs energy of solid solutions would be identifying as many ternary systems as possible of which data can be trusted for NN training. Afterward, a strategy that iteratively improves the performance of NN models by gradually increasing the number of ternary systems needs to be developed. Once a core NN network model is trained with all the known, trustworthy ternary data, it can be anticipated to make reasonable predictions in ternaries where barely any data is available.

4.2. Descriptors beyond elemental compositions and temperature

Although the preliminary results of predicting Gibbs free energies of solid solution phases using a NN trained with a CALPHAD-derived dataset demonstrated certain feasibility, there is a limitation in the applicability of the current approach. It heavily relies on the fidelity of the CALPHAD dataset, and it can be expected to predict reliably where the CALPHAD database was assessed with rich and self-consistent experimental/theoretical data. However, the current approach only uses simple descriptors, i.e., elemental compositions and temperature, to represent Gibbs free energy. Thus, the chemical identities of individual elements are merely given as numerical values without any association with physical/chemical properties. Therefore, NN models will not be able to learn any physical constraints in predicting where the quality of the dataset computed from the CALPHAD database is in question. Also, the current approach cannot make 'true' extrapolations; the boundary conditions of NN models will be confined to elements only when the CALPHAD database is available to provide training data.

Recent studies have shown that introducing relevant physical/chemical descriptors is important to make meaningful predictions with appropriate constraints [10,11]. Using only chemical compositions as descriptors to represent Gibbs free energy of solid solutions was insufficient to capture every atomistic-level phenomenon affecting thermodynamics. Many atomistic-level descriptors have been proposed to predict energetics of thermodynamic phase stability. Some selected examples include the difference in atomic radii between elements, the difference in electronegativity between elements, the valence electron concentration, the mixing enthalpy, the configurational entropy, the $\Omega$ parameter (related to the entropy, enthalpy, and the melting point), the $\Lambda$ parameter (associated with an atom's configuration on a lattice and its radius) and the $\gamma$ parameter (the solid angles of atomic packing for the elements with the largest and smallest atomic sizes), the local electronegativity mismatch between elements, the number of itinerant electrons, and the cohesive energy [27–29]. These atomistic descriptors are related to the intrinsic properties that influence the formation of a solid solution, amorphous phase, and/or intermetallic compound in alloys.

4.3. Binary Gibbs free energies with arbitrary model parameters

The premise of the currently demonstrated approach is that all the binary datasets can be regarded as ground truth, which may be a bold assumption. The CALPHAD community has a convention of using certain numerical tricks to avoid undesirable outcomes in phase diagram calculations. Such examples are 1) using a constant value for the heat capacity of liquid in unary to prevent the Gibbs energy of solid phase from becoming stable again at very high temperatures [30], and 2) using arbitrarily positive values for mixing energy in solid solutions not to have any single-phase field when experimentally reported solubility is extremely small. The second example certainly deteriorates the training of NN surrogate models.

An exemplary case would be a positively evaluated interaction parameter of FCC in Al-Y. The shape and magnitude of mixing energy curves of solid solutions and formation energy convex hull are close to each other for binary alloys with intermetallic phase(s) and exhibit a considerable solid solubility range. Thermodynamic assessments of Al-Ni [31,32] and Al-Cu [33] are such examples. While the convex hull of Al-Y has been negatively evaluated, interaction parameters for FCC and HCP have been assessed to be very high positive numbers, e.g., 90 times of temperature (90$T$) [34]. These arbitrary, largely positive values can successfully prevent showing any single solid phase field of FCC and HCP in binary phase diagram calculations.

However, the error caused by using the numerical convention in the binary will propagate to the higher-order systems. Hence, it may force the use of unnecessary ternary interaction parameters to fix the error of FCC/HCP when Al-Y is combined with other binaries for interpolation. Re-evaluating erroneously evaluated Gibbs free energies in lower-order binaries is daunting as it may disrupt the hierarchical relationships in multi-component CALPHAD databases. Nevertheless, a re-assessment of such constituent binaries is necessary to prevent the propagation of errors to higher-order systems and populate the correct training dataset for the NN approach to properly predict Gibbs free energies in the multi-component systems. A recent assessment of Al-Y modeled the interaction parameter of FCC ($G_{Al,Y:Va}^{fcc}$) as a more realistic -136600 J/mol, which is a start of such efforts [35].

## 5. Conclusion

We employed a data-centric DL approach to overcome the data scarcity and high-dimensionality in Gibbs free energy prediction of ternary solid solution phases. The present study demonstrated that the NN approach has the potential to be useful in diverging the source of acquiring high-fidelity data for CALPHAD assessment, besides experiments and physics-based atomistic simulations. We have used Gibbs free energies of 226 binaries from the state-of-the-art

CALPHAD database and selected ternary energies to augment the binary-only dataset. While predictions from a NN model trained with a binary-only dataset replicated the overall trend of Gibbs free energies in ternary, there is a discrepancy where dominant ternary interaction exists. The other NN surrogate model trained with binary and augmented with selected ternaries showed better accuracy than NN trained with a binary-only dataset. However, the NN prediction accuracy is highly dependent on the choice of augmented ternary data.

Including more ternaries into the augmented dataset is necessary to further improve the prediction accuracy and generalizability of NNs, which requires a more systematic strategy to statistically sample ternary systems in training. In addition, we can consider including elemental level features beyond alloy composition, such as electronegative, atomic radius, and volume, to couple physics and chemistry in the current data analytics approach. Also, a more critical evaluation of assessed binary Gibbs free energies for NN training data is needed, particularly the arbitrarily evaluated ones. It was shown in the present study through several examples that one must carefully select the dataset that goes into the NN training process, which is the same principle that the CALPHAD community emphasizes: critically evaluating the dataset that goes into the CALPHAD model development process. We anticipate that demonstrated DL workflow in the present study could be extended to any multi-component alloy properties predictions.

## Acknowledgment

This research is sponsored by the Artificial Intelligence Initiative as part of the Laboratory Directed Research and Development Program of Oak Ridge National Laboratory, managed by UT-Battelle, LLC, for the US Department of Energy under contract DE-AC05-00OR22725. MLP would like to thank Vladimir Protopopescu and Sam T. Reeves for their suggestions and comments.

DS would like to thank In-Ho Jung for his valuable comments and discussion.

# References

1. C. T. Sims, in *Superalloys 1984 (Fifth International Symposium)* (TMS, 1984), pp. 399–419.
2. N. Saunders and A. P. Miodownik, *CALPHAD (Calculation of Phase Diagrams): A Comprehensive Guide* (Pergamon, Oxford, New York, 1998).
3. Y. A. Chang, S. Chen, F. Zhang, X. Yan, F. Xie, R. Schmid-Fetzer, and W. A. Oates, Progress in Materials Science **49**, 313 (2004).
4. C. Zhang and Y. Yang, MRS Bulletin **47**, (2022).
5. C. J. Bartel, S. L. Millican, A. M. Deml, J. R. Rumptz, W. Tumas, A. W. Weimer, S. Lany, V. Stevanović, C. B. Musgrave, and A. M. Holder, Nature Communications **9**, (2018).
6. J. Yoon, E. Choi, and K. Min, Journal of Physical Chemistry A **125**, 10103 (2021).
7. Y. Mao, H. Yang, Y. Sheng, J. Wang, R. Ouyang, C. Ye, J. Yang, and W. Zhang, ACS Omega **6**, 14533 (2021).
8. S. K. Kauwe, J. Graser, A. Vazquez, and T. D. Sparks, Integrating Materials and Manufacturing Innovation **7**, 43 (2018).
9. C. J. Bartel, A. Trewartha, Q. Wang, A. Dunn, A. Jain, and G. Ceder, Npj Computational Materials **6**, 1 (2020).
10. J. Peng, Y. Yamamoto, J. A. Hawk, E. Lara-Curzio, and D. Shin, Npj Computational Materials **6**, 141 (2020).
11. D. Shin, Y. Yamamoto, M. P. Brady, S. Lee, and J. A. Haynes, Acta Materialia **168**, 321 (2019).
12. J. Peng, N. S. Harsha Gunda, C. A. Bridges, S. Lee, J. Allen Haynes, and D. Shin, Computational Materials Science 111034 (2021).
13. J. Peng, R. Pillai, M. Romedenne, B. A. Pint, G. Muralidharan, J. Allen Haynes, and D. Shin, Npj Materials Degradation **5**, 41 (2021).
14. S. Lee, J. Peng, D. Shin, and Y. S. Choi, Science and Technology of Advanced Materials **20**, 972 (2019).
15. G. L. W. Hart, T. Mueller, C. Toher, and S. Curtarolo, Nature Reviews Materials **0123456789**, (2021).
16. W. Huang, P. Martin, and H. L. Zhuang, Acta Materialia **169**, 225 (2019).
17. M. Ziatdinov, O. Dyck, A. Maksov, X. Li, X. Sang, K. Xiao, R. R. Unocic, R. Vasudevan, S. Jesse, and S. V. Kalinin, ACS Nano **11**, 12742 (2017).
18. S. V. Kalinin, B. G. Sumpter, and R. K. Archibald, Nature Materials **14**, 973 (2015).
19. X. Chong, S.-L. Shang, A. M. Krajewski, J. D. Shimanek, W. Du, Y. Wang, J. Feng, D. Shin, A. M. Beese, and Z.-K. Liu, Journal of Physics: Condensed Matter **33**, 295702 (2021).
20. A. Paszke, S. Gross, F. Massa, A. Lerer, J. Bradbury, G. Chanan, T. Killeen, Z. Lin, N. Gimelshein, L. Antiga, A. Desmaison, A. Köpf, E. Yang, Z. DeVito, M. Raison, A. Tejani, S. Chilamkurthy, B. Steiner, L. Fang, J. Bai, and S. Chintala, Advances in Neural Information Processing Systems **32**, (2019).
21. K. He, X. Zhang, S. Ren, and J. Sun, in *Proceedings of the IEEE Conference on Computer Vision and Pattern Recognition* (2016), pp. 770–778.
22. D. P. Kingma and J. Ba, 3rd International Conference on Learning Representations, ICLR 2015 - Conference Track Proceedings 1 (2014).
23. K. C. Hari Kumar and P. Wollants, Journal of Alloys and Compounds **320**, 189 (2001).
24. D. Shin and Z.-K. Liu, CALPHAD **32**, 74 (2008).
25. D. Shin, A. van de Walle, Y. Wang, and Z.-K. Liu, Physical Review B **76**, 144204 (2007).

26. C. Jiang and B. P. Uberuaga, Physical Review Letters **116**, 105501 (2016).
27. C. Wen, Y. Zhang, C. Wang, D. Xue, Y. Bai, S. Antonov, L. Dai, T. Lookman, and Y. Su, Acta Materialia **170**, 109 (2019).
28. S. Guo, Materials Science and Technology (United Kingdom) **31**, 1223 (2015).
29. L. Ward, A. Agrawal, A. Choudhary, and C. Wolverton, Nature Communications 1 (2015).
30. A. T. Dinsdale, CALPHAD **15**, 317 (1991).
31. I. Ansara, N. Dupin, H. L. Lukas, and B. Sundman, Journal of Alloys and Compounds **247**, 20 (1997).
32. N. Dupin, I. Ansara, and B. Sundman, Calphad: Computer Coupling of Phase Diagrams and Thermochemistry **25**, 279 (2001).
33. S. M. Liang and R. Schmid-Fetzer, Calphad: Computer Coupling of Phase Diagrams and Thermochemistry **51**, 252 (2015).
34. I. Ansara, A. T. Dinsdale, and M. H. Rand, *Thermochemical Database for Light Metal Alloys* (European Cooperation in the Field of Scientific and Technical Research, 1998).
35. J. Huang, B. Yang, H. Chen, and H. Wang, Journal of Phase Equilibria and Diffusion **36**, 357 (2015).

**List of Figures**

Figure 1. A schematic of the current deep learning approach to predict Gibbs free energies of ternary FCC solid solutions

Figure 2. (a) A parity plot that compares the actual data to predictions from an example network, and (b) the loss curves that track the training and validation errors as the number of epochs grows in the training process

Figure 3. Gibbs free energies of (a) Al-Cr-Ni and (b) Al-Cu-Fe isopleths at three iso-concentrations predicted from NN (circles) and computed from the TCHEA5 CALPHAD database (lines) at six different temperatures, 300, 500, 700, 900, 1100, and 1300K.

Figure 4. Comparison between Gibbs free energies of (a) Al-Cu-Ni and (b) Co-Cu-Mn at 300K computed from the CALPHAD database (TCHEA5) and NN predicted values.

Figure 5. (a) Comparison between binary interpolated and assessed Co-Cu-Mn at 300K, and (b) isopleths at three iso-concentrations at 700K and 900K.

Figure 6. Different numbers of isopleths (9, 24, 49, and 99) of Co-Cu-Mn at 300K from the CALPHAD database (TCHEA5).

Figure 7. The accuracy of NN, represented with the coefficient of determination ($R^2$) as a function of the amount of augmented ternary data. The data set with zero percent of ternary data correspond to the binary-only training set.

Figure 8. NN predicted Gibbs free energies of Al-Cr-Ni at 850K trained with binary-only and ternary augmented dataset. The errors are evaluated with respect to the ground truth, computed from the TCHEA5 database.

Figure 9. NN predicted Gibbs free energies of Co-Cu-Mn at 850K trained with binary-only and ternary augmented dataset. The errors are evaluated with respect to the ground truth, computed from the TCHEA5 database.

Figure 10. NN predicted Gibbs free energies of Al-Cu-Mn at 850K trained with binary-only and ternary augmented dataset. The errors are evaluated with respect to the ground truth, computed from the TCHEA5 database.

Figure 11. NN predicted Gibbs free energies of Al-Fe-Mn at 850K trained with binary-only and ternary augmented dataset. The errors are evaluated with respect to the ground truth, computed from the TCHEA5 database.

Figure 12. NN predicted Gibbs free energies of Al-Cu-Zn at 850K trained with binary-only and ternary augmented dataset. The errors are evaluated with respect to the ground truth, computed from the TCHEA5 database.

Figure 13. NN predicted Gibbs free energies of Al-Mo-Ni at 850K trained with binary-only and ternary augmented dataset. The errors are evaluated with respect to the ground truth, computed from the TCHEA5 database.